\documentclass[11pt]{article}

\usepackage{geometry} 
\usepackage{graphicx}
\usepackage[breaklinks]{hyperref}
\usepackage{natbib}

\title{Tracing a relativistic Milky Way within the RAMOD measurement protocol}

\begin{document}

\maketitle

\begin{center}
\author{ Mariateresa \, Crosta } \\
Osservatorio Astronomico di Torino -- Istituto Nazionale di Astrofisica
\\ via Osservatorio, 20 -- 10025, Pino Torinese, Italy\\ 
crosta@oato.inaf.it
\end{center}

\begin{abstract}

Advancement in astronomical observations and technical instrumentation implies taking into account the general relativistic effects due the  gravitational fields encountered by the light while propagating from the star to the observer. Therefore, data exploitation for Gaia-like space astrometric mission (ESA, launch 2013) requires a fully relativistic interpretation of the inverse ray-tracing problem, namely the development of a highly accurate astrometric models in accordance with the geometrical environment affecting light propagation itself and the precepts of the theory of measurement. This could open a new rendition of the stellar distances and proper motions, or even an alternative detection perspective of many subtle relativistic effects suffered by light while it is propagating and subsequently recorded in the physical measurements.

\end{abstract}

\section{Introduction}
The role of astrometry has been revitalized thanks to the space mission Gaia \cite{gaia} which will be launched by ESA not earlier than September 2013.
The expected end-of-life astrometric performance, at the level of $\mu$as accuracy, requires to take into account light deflections effects due to the Solar System bodies. This implies that any astrometric measurement has to be modelled in a way that stellar light propagation and detection should be both conceived in a general relativistic framework. As matter of fact, the trajectory of a photon  is traced by solving the null geodesic in a curved space-time dictated by General Relativity (GR) and at the same time, the detection process usually takes place in a geometrical environment  generated by a n-body distribution as it is that of our Solar System (SS). 

Nowadays, a few approaches exist that model light propagation in a relativistic context. Among them, the post-Newtonian (pN) and the post-Minkowskian (pM) approximations are those mainly used (\cite{kopetma, klio, teyetlpl} and references therein).  Inside the Consortium  constitued for the Gaia data reduction (Gaia CU3, Core Processing, DPAC) two different formulations of relativistic light propagation have been developed to model astrometric observations, of distant sources by an SS observer: (i) the GREM formulation \cite{klio}, known as the Gaia baseline model IAU coordinate based \cite{soffetal}, and (ii) the RAMOD model (\cite{fdfetmtc}, 2006), an alternative approach fully compliant with the precepts of  local measurement in a relativistic setting. Actually RAMOD is a family of astrometric models of increasing intrinsic accuracy conceived to solve the inverse ray-tracing problem in a general relativistic framework. 
Their theoretical equivalence to the 1-$\mu$as accuracy level has been recently demonstrated (\cite{Crosta2011}  and reference therein) and will be exploited, in a process, called in the Gaia jargon, Astrometric Verification Unit (AVU) by comparing the results of two fully independent astrometric reconstructions of the celestial sphere to assess all-sky scientific reliability on position, including parallax, and proper motions.  The link between the models is crucial as far as the Gaia's goal is concerned: the unbiased measurements, i.e., independently from models, the most fundamental astrophysical stellar parameters (absolute distance, angular position, velocity, and mass) for approximately 1 billion individual stars.

It can be inferred that the treatment of light propagation in time-dependent gravitational fields encompasses issues from fundamental astronomy to cosmology (\cite{will}, \cite{tury}, \cite{mtcetmig}, \cite{kopetgwin}, \cite{damour}, \cite{fdfetpre} and references therein). The accurate measurement of the motions of stars in our Galaxy can also provide access to the cosmological signatures in the disk and halo, while astrometric experiments from within our Solar System can unequally probe possible deviations from GR just one century after Einsteins's great discoveries. 
With the Gaia mission approaching launch, Relativistic Astrometry is about to trace the geometry of the visible Milky Way.

\section{\label{sec:sec1}The astrometric problem}

The astrometric problem consists, firstly, in solving the null geodesic for the single stellar photon,  in order to trace back the light trajectory to the initial position of the emitting source and, then, determine its astrometric parameters through the astrometric observable, according to the chosen reference frames. Differently from the other approaches, RAMOD's full solution requires the integration of a set of coupled non-linear differential equations, called ``master equations''. The unknown of these equations is the local line-of-sight {\boldmath$\bar\ell$} as measured by the fiducial observer $\mathbf {u}$ at the point of observation in her/his rest-space. At the time of observation,  {\boldmath$\bar{\ell}$} provides the boundary condition for uniquely solving the light path by means of the relativistic definition of the observable (\cite{mtcetvec}) and the satellite-observer frames (\cite{binietmtc}). The main purpose of the RAMOD approach is to express the null geodesic through all the physical quantities entering the process of measurement without any approximations, in order to entangle all the possible interactions of light with the background geometry. 
RAMOD uses a 3+1 characterization of space-time in order to measure physical phenomena along the proper time and on the rest-space of a set of fiducial observers according to the following measurement protocol (\cite{fdfetbini}): i) specify the phenomenon under investigation; ii)
identify the covariant equations which describe it; iii) identify the observer who makes the measurements; iv) chose a frame adapted to that observer allowing the space-time splitting into  the observerÕs space and time; vi) understand the locality properties of the measurement under consideration (local or non-local with respect to the background curvature); vii) identify the frame components of those quantities which are the observational targets; viii) find a physical interpretation of the above components following a suitable criterium; ix) verify the degree of the residual ambiguity in the interpretation of the measurements and decide the strategy to evaluate it (i.e. comparing what already is known).

 Solving the astrometric  problem in practice means to compile an astrometric catalog  with the same order of accuracy as the measurements. To what extent, then, is the process of star coordinate ÒreconstructionÓ consistent with General Relativity \&Theory of Measurements?

\section{\label{sec:sec3}The static solution of the astrometric problem}

Gaia-like measurement takes place inside the Solar System, i.e. a  weakly relativistic gravitationally bound system, described by the metric $g_{\alpha \beta }= \eta_{\alpha \beta} + h_{\alpha \beta} + O(h^2)$. Now, in order to gauge how much curvature can be considered local or not with respect to the measurement,  let us resort the virial theorem which requires an energy balance of the order of $|h_{\alpha \beta}|\le U/c^{2} \sim v^{2}/{c}^{2}$, where $v$ is the characteristic relative velocity within the system~\footnote{For a typical velocity  $\sim 30$ km/s, $(v/c)^2 \sim$  1 milli-arcsec}. Therefore the level of accuracy is fixed by the order of the small quantity $\epsilon\sim (v/c)$. Since the system is weakly relativistic, the perturbation tensor $h_{\alpha \beta}$ contributes with even terms in $\epsilon$ to $g_{ 00 }$ and $g_{ ij}$ (lowest order $\epsilon^{2}$) and with odd terms in $\epsilon$ to $g_{0i}$ (lowest order $\epsilon^{3}$, \cite{gravitation,fdfetbini}); its spatial variations are of the order of $|h_{\alpha\beta}|$, while its time variation is of the order of $\epsilon| h_{\alpha\beta}|$. 
This means that at the order of $\epsilon^3$, not only the time dependence of the background metric cannot be ignored any longer, but also the vorticity, which measures -in the process of foliation- how a world-line of an observer rotates around a neighboring one, can be neglected being proportional to the $g_{0 i}$ term of the metric (see details in \cite{Crosta2011}). Consequently, it is not possible to define a rest-space of a fiducial observer that covers the entire space-time.
Any observer $\mathbf {u}$ can be considered at rest with respect to the coordinates $x^i$ {\em only locally}, and for this reason $\mathbf u$ is called the {\em local barycentric observer}, as identified in \cite{fdfetvec}. 
The master equations satisfied by the vector field {\boldmath$\bar\ell$} up to the $\epsilon^3$ order of accuracy are
\begin{eqnarray}
\frac{d\bar\ell^0}{d\sigma}&= &\bar\ell^i\bar\ell^j h_{0j,i}+\frac{1}{2}  h_{00,0} \label{eq:diffeq0}\\
&{}&\nonumber\\
\frac{d\bar\ell^k}{d\sigma}&=&\frac{1}{2} \bar\ell^k\bar\ell^i\bar\ell^j h_{ij,0}-\bar\ell^i\bar\ell^j\left( h_{kj,i}-\frac{1}{2} h_{ij,k}\right) \nonumber \\
&-& \frac{1}{2}\bar\ell^k\bar\ell^i  h_{00,i}- \bar\ell^i\left(h_{k0,i}+h_{ki,0} - h_{0i,k}\right) \nonumber \\
&+&\frac{1}{2} h_{00,k} +\bar \ell^{k} \bar \ell^{i} h_{0i, 0}, \label{eq:diffeqk}
\end{eqnarray}
named ``RAMOD4 master equations''  in the dynamical case (\cite{fdfetvec, Crosta2011}), being $\sigma$ the parameter of the null geodesic.
Note that there is a differential equation also for the $\bar\ell^{0}$ component, which represents an opportunity to better decipher light propagation in future developments. 

The $\epsilon^2$ regime, instead, is referred as the ``static case'', or ``static space-time'', i.e. a stationary space-time in which a time-like Killing vector field ${\mathbf u}$ has vanishing vorticity (\cite{fdfetmtc}). In this case the parameter $\sigma$ on ${\mathbf u}$  is the proper time of the physical observers who transport the spatial coordinates without shift.   Any hypersurface $t(x,y,z)=constant$, at each different coordinate time $t $, can be considered the rest space {\em everywhere} of the observer ${\mathbf u}$ and the geometry that each photon feels is, then, identified with the weak relativistic metric where $g_{0i}=0$.
In these circumstances we can define a one-parameter local diffeomorphism which maps each point of the null geodesic to the point on the slice at the time of observation, say $S(t_o)$ (\cite{fdfetmtc}):
\begin{eqnarray}
\frac{d\bar{\ell}^{k}}{d\sigma}&=& - \bar{\ell}^{k} \left(\frac{1}{2}
\bar{\ell}^{i}h_{00,i}\right) - \delta^{k s} \left( h_{s j, i} 
-\frac{1}{2} h_{ij, s}\right)\bar{\ell}^{i}\bar{\ell}^{j} \nonumber \\
&+& \frac{1}{2}\delta^{ks} h_{00, s}. \label{eq:geodint}
\end{eqnarray}
Equations (\ref{eq:geodint}) determine light propagation in the static case, and are called ``RAMOD3 master equations''  (\cite{fdfetmtc}).
 This equation can be solved  analytically.  Let us assume a weak field metric such as:
\begin{eqnarray}
 h_{00} \equiv h & \simeq & \sum_{a}  h_{(a)} =\sum_{a}  \frac{2G\mathcal{M}_{(a)}}{c^{2}r_{(a)}}  \nonumber \\
h_{ij} & \simeq & h\,\delta_{ij},\nonumber 
\end{eqnarray}
where the index $a$ is refereed to the gravitational sources. Equations  (\ref{eq:geodint}) can be reduced by considering that  $\bar{\ell}^{i}\bar{\ell}^{j}\delta_{ij}=1+\mathcal{O}\left(h^{2}\right)$, namely:
\begin{equation}
\frac{\mathrm{d}\bar{\ell}^{k}}{\mathrm{d}\sigma} + \frac{3}{2}\bar{\ell}^{k}(\bar{\ell}^{i}h_{,i})- h_{,k}+\mathcal{O}(h^{2}) =0 . \label{eq:master-equation-2}
\end{equation}
Only a vorticity-free space-time allows to parametrize simultaneously the mapped trajectory with respect to the Center of Mass (CM) of the gravitational bodies on $S(t_o)$; if the Euclidean scalar product is applied, the RAMOD procedure for the parametrization generalizes the one used in \cite{kopetma} or  \cite{klio} (\cite{Crosta2011}). 
Then, within the scale of a vorticity-free geometry,  we can always express the mapped trajectory in such a parametrized form: 
\begin{equation}
x^i= \hat{\xi^i}+ \int_o^{\hat\tau} \bar{\ell}^i d\hat{\tau},
\label{eq:parametrization}
\end {equation}
where $\hat{\xi}^i$ is the impact parameter w.r.t. CM,  $\hat \tau = \sigma - \hat \sigma$, being $\hat \sigma$ the value of geodesic parameter at the point of the closest approach.  
If we approximate $\bar{\ell}^i$ in function of small perturbations with respect to the unperturbed light direction $\bar \ell_{\not {0}}$, i.e.:
 \begin{equation}
 \bar{\ell}^i = \bar{\ell}^{i}_{\not 0} + \delta \bar{\ell}^{i} + (\delta \bar{\ell}^{i})^2 + ...
\label{eq:lo-exp}
\end {equation}
we note that term of the order of  $(\delta \bar{\ell}^{i}) $ can be neglected in the master equations because of the order of $O(h)$.

This implies that also equation (\ref{eq:parametrization}), for our purpose, can be approximated as:
\begin{equation}
x^i= \hat{\xi^i}+  (\bar{\ell}^i_{\not 0} +\delta \bar\ell^i) \hat{\tau} + O\left(h^2 \right) ,
\label{eq:parametrization-2}
\end {equation}
where $\delta \bar\ell^j$  is of the order of the deflection, i.e. $\epsilon^2$.

After these assumptions, we reformulate the relative distance between the photon and source coordinates, $r^{i}_{(a)} $, as
\begin{equation}
 \hat r^i (\hat \tau) =\hat{\xi}^i+ (\bar{\ell}^i_{\not 0}+ \delta \bar\ell^k ) \hat\tau  - {x}_{(a)}^i=   \hat r_p + (\bar{\ell}^i_{\not 0}+ \delta \bar\ell^k )  \hat\tau ,
\label{eq:rjo}
\end{equation}
 where 
\begin{equation}
 \hat r_p^i = \hat{\xi}^i -{x}^i_{(a)},
\end{equation}
is the relative distance between the source coordinates and the photon impact parameter w.r.t. the CM.   

By using the scalar and vectorial products  the impact parameter with respect to the source at the point of the closest approach can be defined as:
\begin{equation}
d(\hat \tau)^k= [\bar \ell \times (\hat r(\hat \tau) \times \bar \ell)]^k .
\label{eq:planet-impact-parameter}
\end{equation}
Note, it does not depends on $\hat \tau$ by definition, so it is always
\begin{equation}
d^k_p= [\bar \ell \times (\hat r_p \times \bar \ell)]^k ,
\label{eq:planet-impact-parameter}
\end{equation}

Then equation (\ref{eq:master-equation-2}) is simplified as
\begin{eqnarray}
\mathrm{\Delta}\bar{\ell}^{k}&=&\frac{2G}{c^{2}} \sum_{a} \mathcal{M} 
\left \{ 
\int_{\hat \tau}^{\hat \tau_o}   \left [\frac{1}{2} \bar{\ell}^{k}  (\vec{\bar{\ell}} \cdot   \vec{\hat r}) -  \hat{d}^k  \right]  \frac{d \hat \tau }{\hat r^3} 
\right \} + O \left( h^2 \right),
\label{eq:red2-master-equation-qbodya}
\end{eqnarray}
which is easly solved as
\begin{eqnarray}
\mathrm{\Delta}\bar{\ell}^{k}&=&\frac{2G}{c^{2}} \sum_{a} \mathcal{M} \frac{1}{d_p^2} 
\left \{ 
  \left [ \frac{1}{2} \bar{\ell_{\not 0}}^{k}  ({\bar{\ell_{\not 0}}} \cdot  {r}_p) - d_p^k   \right]   \left[  \frac{ ({\bar\ell}_{\not 0} \cdot  {\hat r})  }  {  \hat r} \right]^{\hat \tau_o}_{\hat \tau}  -   \frac{1}{2} \bar{\ell_{\not 0}}^{k} \left[  \frac{({\hat r} \cdot   {r}_{p}) }{ \hat r}\right]^{\hat \tau_o}_{\hat \tau} \right \}\nonumber \\
&+& O \left( h^2 \right)
\label{eq:master-equation-static-monopole}
\end{eqnarray}
namely
\begin{eqnarray}
\mathrm{\Delta}\bar{\ell}^{k}&=&\frac{2G}{c^{2}} \sum_{a} \mathcal{M} 
\left \{ - \frac{\bar{\ell_{\not 0}}^{k} }{2 \hat r(\hat \tau_o)} 
  +  \frac{\bar{\ell_{\not 0}}^{k} }{  2 \hat r } -  \frac{d_p^k}{d_p^2}  \left[ \frac{{\bar \ell}_{\not 0} \cdot  {\hat r} }{\hat r}  -  \frac{ \vec{\bar \ell}_{\not 0} \cdot {\hat r}(\hat \tau_o)}  {  \hat r(\hat \tau_o) } \right]  \right \} \nonumber \\
 &+&O \left( h^2 \right).
\label{eq:red9-master-equation-n-body}
\end{eqnarray}

This last formula is easily converted in that one found by Klioner (2003), if we consider a source at infinity ($\bar \ell_{\not 0}^k \equiv  \sigma^k$):
\begin{eqnarray}
\Delta \bar \ell^k &\approx &\frac{2G}{c^{2}} \sum_{a} \mathcal{M} 
\left \{ -  \frac{\sigma^{k} }{ 2\hat r(\tau_o) } - \frac{d_p^k}{ d_p^2}  \left[ 1  + \frac{ {\sigma}\cdot  
{\hat r}(\tau_o)}  {  \hat r(\tau_o)} \right] 
 \right \} +O \left( h^2 \right).
\label{eq:klo}
\end{eqnarray}
Then, if the observable shift w.r.t. to the direction at infinity is  $ \delta \sigma = c^{-1} [\sigma \times \Delta \dot x \times \sigma ]^k$, one gets
\begin{eqnarray}
\delta \sigma^k &\approx &\frac{2G}{c^{2}} \sum_{a} \mathcal{M} 
\left \{ -  \frac{d_o^k}{ d_o^2}  \left[ 1  +  \frac{ {\sigma}\cdot  
{\hat r}(\tau_o)}  {  \hat r(\tau_o)}   \right] 
 \right \} +O \left( h^2 \right).
\label{eq:klo-def}
\end{eqnarray}

As far as RAMOD method is concerned, instead, we need to insert $\bar \ell^k (\tau_o)$, {\it i.e.} the {\it observed value} of the tangent vector to the light ray, into the astrometric observable, i.e. the cosine \cite{mtcetvec}. 
Then, we need to integrate again (\ref{eq:red9-master-equation-n-body})
\begin{eqnarray}
\bar{\ell}^{k} (\hat \tau_o) \Delta \hat \tau- \int^o_*  \frac{dx^k}{d\hat \tau} d\hat \tau&=&\frac{2G}{c^{2}} \sum_{a} \mathcal{M} 
\int_*^o \left \{ - \frac{\bar{\ell_{\not 0}}^{k} }{2 \hat r(\hat \tau_o)} 
  +  \frac{\bar{\ell_{\not 0}}^{k} }{  2 \hat r} -    \frac{d_p^k}{d_p^2} \left[ \frac{({\bar \ell}_{\not 0} \cdot  \hat r_p ) + \hat \tau_o}{\hat r(\hat \tau_o)}  \nonumber \right.\right. \\
 &&\left. \left. -  \frac{ ({\bar \ell}_{\not 0} \cdot  { \hat r})}  {  \hat r } \right] 
 \right \} d\hat \tau +  O \left( h^2 \right).
\label{eq:red11-master-equation-n-body}
\end{eqnarray}
where $\Delta \hat\tau= \hat \tau_o - \hat \tau_*$. The expression of $\Delta \tau $ as flight time from the star to the observer - measured with respect to the local barycentric observer-  is given by Bertone et al. in the comparison between RAMOD and the Time Transfer Function method  \footnote{ Bertone, S. Minazzoli, O. Crosta, M. Le Poncin-Lafitte, C. Vecchiato, A., and Angonin, M.C. {\it Time Transfer functions as a way to validate light propagation solutions for space astrometry}, submitted to Classical and Quantum Gravity} (\cite{teyetlpl}). 
Finally, one can straightforwardly check that the barycentric coordinate of the star assumes the following perturbed form 
\begin{eqnarray}
x^k_*&=&  x^k_o + \bar{\ell}^{k}_o \Delta \hat \tau + \frac{2G}{c^{2}} \sum_{a} \mathcal{M} 
\left \{ \frac{\bar \ell^k_{\not 0}}{2} Log \left[\frac{({\bar \ell}_{\not 0} \cdot  \hat r_o) + \hat r_o }{({\bar \ell}_{\not 0} \cdot \hat r_*) + \hat r_*} \right]+ \right.  \nonumber \\
 & &  \left. \left[\frac{\bar \ell^k_{\not 0}}{2 \hat r_o } +  \frac{d^k_p (\bar \ell_{\not 0} \cdot \hat r_o) }{\hat r_o d^2_p } \right] \Delta \hat \tau + \frac{d^k_p }{d^2_p}  (\hat r_o - \hat r_*)  \right\}+  O \left( h^2 \right).
\label{eq:red11-master-equation-n-body}
\end{eqnarray}

\section{\label{sec3}Matching physics and coordinates at high accuracy}
The fact that light tracing is different with or without the vorticity term make evident how the RAMOD recipe, based on a measurement protocol, differs from a direct "coordinate" approach which, instead, does not need to discriminate the accuracy of the geometry to be involved. 
The quantity $\mathbf{\bar{\ell}}$ is the unitary four-vector representing the \emph{local line-of-sight} of the incoming photon as measured by the local observer $\mathbf{u}$ in his/her gravitational environment; it represents a physical quantity in any case, with or without vorticity. By implementing its coordinate expressions straightway, equations (\ref{eq:diffeqk}), i.e. those for the spatial components, are converted into the coordinate ones derived in \cite{kopetma}  at the first pM approximation of the null geodesic (\cite{Crosta2011}). This result was expected, since both models are deduced from the null geodesic in a weak field regime. Then, once such an equivalence is obtained, one could solve  the master equation in the RAMOD framework by applying the same procedure adopted in \cite{kopetma}. However, consistently with the reasoning of the previous section, only RAMOD3 master equation can be transformed into the solution given by  \cite{kopetma}, since the parametrization in RAMOD is possible only in a vorticity-free space-time where admits also an analytical solution.  In fact, if one assumes a perturbed straight line trajectory,  the equivalence of the two parametrizations implies a change of coordinates which transforms equation (\ref{eq:diffeqk}) into the same parametrized equation (36) used in \cite{kopetma}. Nevertherless, the integration of the null geodesic in \cite{kopetma} intends to consider the gravitomagnetic effects. In addition,  the metric coefficients $h_{\alpha\beta}$ depend on the retarded distance $r_{(a)}$ as discussed in \cite{fdfetvec}. This means that one has to compute the spatial coordinate distance $r_{(a)}$ from the points on the photon trajectory to the {\em a}-th gravity source at the appropriate retarded time and up to the required accuracy. 
Hence, if we wish our model be accurate to $\epsilon^3$, it suffices that the retarded distance $r$ contributes to the gravitational potentials, which we remind are at the lowest of order $\epsilon^2$, with terms of the order of $\epsilon$. 
Instead, to the order of $\epsilon^2$ (static geometry), the contribution of the relative velocities of the gravitating sources can be neglected. Indeed, one can choose to further expand the retarded distance in order to keep the terms depending on the source's velocity up to the desired accuracy. Obviously the effects due to the bodies' velocity cannot be related to a gravity-magnetic effect, at least up to the scale where the vorticity can be neglected.  Actually, the positions of the bodies can be recorded as subsequent snapshots onto the mapped trajectories and deduced as "postponed" corrections in the reconstruction of the photon's path.  

The importance of the measurement protocol in setting the correct role of the coordinates, and thus avoiding misinterpretations of parallel but different quantities, is also discussed in \cite{mtcetvec}, where, within the context of the Gaia mission (ESA, \cite{gaia}), a first comparison between RAMOD and GREM (Gaia RElativistic Model, \cite{klio}) was carried out via the extrapolation of the aberrational term in the {\it local} light direction. Differences, that already exist at the level of the aberration effect, suggest particular care in the interpretation of the final catalog. Another example which shows how the accurate inclusion of the geometry redraws a standard measurement,  is given by the formula for the Doppler shift  in \cite{fdfetpre}. The spectroscopic and astrometric data that will be provided by the new generation of satellites can be implemented with one another, thus leading to a general-relativistic Doppler which is exact up to and including the $\epsilon^3$ terms. It is also showed that a previously proposed Doppler-shift formula is definitely not adequate to this task, since it misses relevant relativistic corrections already at  $\epsilon^2$.

\section{\label{sec:comments}Conclusions}

Modeling light propagation is intrinsically connected to the identification of the geometry where photons move. 
The different conception of RAMOD provides a method to exploit high accurate observations to their full extent, as it could be the case for the astrometric data coming from the ESA mission Gaia, possibly a new beginning in the field of Relativistic Astrometry. 
The comparison between different light modeling approaches is extremely important since Gaia will ÒchangeÓ our scientific vision and we are implementing new methods using real data.  
By comparing different formulations of a null geodesic we have the opportunity the exploit the advantages of the different methods and improve on our understanding of light propagation.  
As far as RAMOD method is concerned, the geometrical distinction between the master equations  introduces a criterium to disentangle coordinate and physical effects.

In RAMOD the vorticity term cannot be neglected at the order of $\epsilon^{3}$: ignoring it locally is valid only in a small neighborhood compared to the scale of vorticity itself. When the vorticity term is needed the light trajectory cannot be laid out on a unique rest-space of simultaneity from the observer to the star, wherever the latter could be located. Without vorticity RAMOD allows a parametrization of the light trajectory and sets the level of reciprocal consistency with the existing approaches. 
Only master equations of RAMOD4, i.e. the case of a dynamical space-time, fully preserves the active content of gravity.  This solution is enough accurate for implementing Relativistic Astrometry beyond Gaia. 
Considering the number of objects that can be observed in high accuracy regime, the local line-of-sight, as a physical entity, can be also used in the future for an Òinverse parameter problemÓ approach, able to statistically determine the metric also outside the Solar System (\cite{tara}).

\bibliography{ae100prg_crosta}
\bibliographystyle{plainnat}

\end{document}